\documentclass[letter,seceq,twocolumn]{jpsj3}
\usepackage[usenames]{color}
\def\v#1{\mib #1}

\def\Ms{M_{\rm s}}
\newcommand{\bra}[1]{\left\langle {#1} \right\vert}
\newcommand{\ket}[1]{\left\vert {#1} \right\rangle}

\def\enu{n}
\def\MAFH{MAFH}
\def\Ms{M_{\rm s}}
\def\Nc{N_{\rm c}}
\def\Etot{E^{\rm tot}_{\rm G}}
\def\EHal{{\tilde{E}}}
\def\EG{{E_{\rm G}}}
\def\epsg{{\epsilon_{\rm G}}}
\def\epsgtilde{{\tilde{\epsilon}_{\rm G}}}

\title
{
Ground-State Phase Diagram of $S=1$ Diamond Chains
}

\author
{
Kazuo {Hida}$^{1}$\thanks{E-mail: hida@mail.saitama-u.ac.jp} and Ken'ichi {Takano}$^{2}$
}

\inst
{$^{1}$Division of Material Science, Graduate School of Science and Engineering, \\ Saitama University, Saitama, Saitama 338-8570, Japan\\ 
$^{2}$Toyota Technological Institute, 
Tenpaku-ku, Nagoya 468-8511, Japan}

\recdate
{December 2, 2016
}

\abst
{
We investigate the ground-state phase diagram of a spin-1 diamond chain. 
Owing to a series of conservation laws, any eigenstate of this system can be expressed using the  
eigenstates of finite odd-length chains or {infinite chains} with spins 1 and 2. 
The ground state undergoes quantum phase transitions 
with varying $\lambda$, a parameter that  controls frustration.  {Exact upper and lower bounds for the phase boundaries between these phases are {obtained}. The phase boundaries are determined numerically in the region not explored in a previous work [Takano {\it et al.}, J. Phys.: Condens. Matter {\bf 8} (1996) 6405].}
}

\begin{document}

\sloppy

\maketitle

\begin{figure} 
\centerline{\includegraphics[width=5cm]{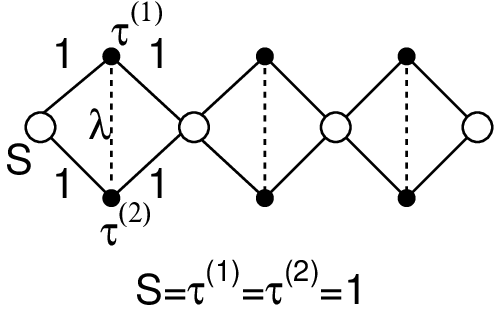}}
\caption{Structure of the diamond chain.}
\label{lattice}
\end{figure}

The quantum effects in frustrated magnets have been extensively studied in condensed matter physics.\cite{intfrust} The interplay of quantum fluctuation and frustration gave birth to various  exotic quantum phases. 
Among a variety of models and materials with strong frustration, the diamond chain, which consists of successive diamond-shaped units as depicted in Fig. \ref{lattice}, has been attracting the interest of many condensed matter physicists. 
{In addition to the well-known natural mineral azurite\cite{kiku2,kiku3} which is  
{a spin-1/2} distorted diamond chain, {it has been pointed out quite recently that  
 [Ni$_3$(OH)$_2$(O$_2$C-C$_2$H$_2$-CO$_2$)(H$_2$O)$_4$]E2H$_2$O\cite{guillou} can be regarded as a spin-1 distorted
 diamond chain.\cite{kuni,kiku1}}} 

From the theoretical viewpoint, the diamond chain {is remarkable, since it} has been rigorously treated to some extent in the absence of distortion.\cite{tks} 
{The ground-state phases of the spin-1/2 diamond chain are fully understood.} The distorted version of this model with spin 1/2 has also been intensively investigated theoretically\cite{ottk,otk} as a model for  azurite. 
 The case of mixed spin diamond chain with $S=1/2$ and 1, which has a Haldane-type ground state in the absence of frustration, has also been extensively studied theoretically\cite{nig1,nig2,tsh,hts1,hts2,hts3,htanis}.
 
 The case of higher spin uniform diamond chain has been briefly discussed in Ref. \citen{tks}. With the increase in spin magnitude,  much richer phase diagrams than the case of spin-1/2 are expected.  However, the full phase diagram has not yet been obtained even for the case of $S=1$.  
We expect that the discovery of the spin-1 distorted
 diamond chain material\cite{kiku1} opens a new field in the physics of frustrated quantum magnetism: 
the ground states of ideal spin-1 diamond chains not only have richer phases compared with the spin-1/2 case but also can generate an even larger variety of phases in the presence of realistic perturbations such as lattice distortion, anisotropy, and interchain coupling.  
Motivated by this expectation,  
we determine the complete ground-state phase diagram of the ideal spin-1 diamond chain based on the exact wave function in the present work.

The Hamiltonian of the $S=1$ diamond chain 
in Fig.~\ref{lattice} is represented as 
\begin{align}
{\cal H} &=\sum_{l=1}^{L} \Big[\v{S}_{l}\v{\tau}^{(1)}_{l}+\v{S}_{l}\v{\tau}^{(2)}_{l}+\v{\tau}^{(1)}_{l}\v{S}_{l+1}+\v{\tau}^{(2)}_{l}\v{S}_{l+1}\nonumber\\
& + \lambda\v{\tau}^{(1)}_{l}\v{\tau}^{(2)}_{l}\Big], \label{ham}
\end{align}
where $\v{S}_{l}$ (hereafter called $S$-spin) and $\v{\tau}^{(\alpha)}_{l}\ (\alpha=1,2)$ (called $\tau$-spin) are spin operators with magnitude 1 in the $l$th unit cell.  
The bonds connecting the spins $\v{\tau}^{(1)}_{l}$ and $\v{\tau}^{(2)}_{l}$ are called diagonal bonds.

By defining the composite spin (called  $T$-spin) 
$\v{T}_l \equiv \v{\tau}^{(1)}_{l}+\v{\tau}^{(2)}_{l}$ 
for all $l$, the Hamiltonian (\ref{ham}) is also expressed as
\begin{align}
{\cal H} &=\sum_{l=1}^{L} \left[\v{S}_{l}\v{T}_{l}+\v{T}_{l}\v{S}_{l+1}+ \frac{\lambda}{2}\left(\v{T}^2_{l}-4\right)\right]  \label{ham1}.
\end{align}
In this representation, it is clear that $\v{T}_l^2$ ($l=1,2,..., L$) commutes with ${\cal H}$. Hence, $T_l$ defined by $\v{T}_l^2 = T_l(T_l + 1)$ is a good quantum number that takes the values 0, 1, and 2.  Thus, within the sector specified by the sequence of $T$-spin magnitudes $\{T_l\}$, the Hamiltonian (\ref{ham1}) is equivalent to the following Hamiltonian
\begin{align} 
{\cal H}(\{T_l\},\lambda) &=\tilde{\cal H}(\{T_l\}) + \frac{\lambda}{2}\sum_{l=1}^{L}(T_{l}+1)T_l-{2L\lambda}, \label{ham2}
\end{align}
where $\tilde{\cal H}(\{T_l\})$ is the Hamiltonian of the mixed spin antiferromagnetic Heisenberg chain (\MAFH) defined by
\begin{align}
\tilde{\cal H}(\{T_l\})&=\sum_{l=1}^{L} \left[\v{S}_{l}\v{T}_{l}+\v{T}_{l}\v{S}_{l+1}\right] \label{ham3}
\end{align}
with fixed $T$-spin magnitudes $\{T_l\}$.

If $T_l=0$, the $\tau$-spins on the $l$th diagonal bond form a singlet pair, which we call a dimer. 
The dimer breaks the correlation between the $S$-spins on both sides. 
The ground state of the finite chain consisting of 
$S$-spins and $\tau$-spins in the region between two dimers is called a cluster-$n$, if the region includes $n$ diagonal bonds and 
all $T$-spins on them are nonzero. 
A cluster-$n$ is specified by an $n$-membered subsequence $\{T_l\}_{i,n}\equiv \{T_i \cdots T_{i+n-1}\}$  of $\{T_l\}$ with $T_l=1$ or 2. 
The cluster-$n$  with the sequence $\{T_l\}_{i,n}$ is the ground state of the Hamiltonian
\begin{align}
{\cal H}(\{T_l\}_{i,n}) 
&=\tilde{\cal H}(\{T_l\}_{i,n})
+ \frac{\lambda}{2} \sum_{l=i}^{i+\enu-1} \left[(T_{l}+1)T_l-4\right] 
\label{eq:cln}
\end{align}
with 
\begin{align}
\tilde{\cal H}(\{T_l\}_{i,n})&=\sum_{l=i}^{i+n-1} \left[\v{S}_{l}\v{T}_{l}+\v{T}_{l}\v{S}_{l+1}\right], 
\label{eq:finite_mafn}
\end{align}
which is the Hamiltonian for the {\MAFH} with $2n+1$ spins and the open boundary condition. 
We denote the lowest eigenvalues of (\ref{eq:cln}) and (\ref{eq:finite_mafn})  by $\EG(\{T_l\}_{i,n},\lambda)$ and $\EHal_{\rm G}(\{T_l\}_{i,n})$, respectively. 
They satisfy the relation  
\begin{align}
&\EG(\{T_l\}_{i,n},\lambda)=\EHal_{\rm G}(\{T_l\}_{i,n}) 
+ \frac{\lambda}{2}\sum_{l=i}^{i+n-1}[(T_{l}+1)T_l-4]. 
\end{align}

The lowest eigenstate of the total Hamiltonian (\ref{ham2}) with fixed $\{T_l\}$ is a tensor product of dimers at zero $T$-spins and of {cluster-$n$'s} between them. 
If the state includes $\Nc$ cluster-$n$'s, the lowest eigenvalue is written as 
\begin{align}
\Etot(\{T_l\},\lambda) = 
\sum_{\alpha=1}^{\Nc} \EG(\{T_l\}_{i_{\alpha},n_{\alpha}},\lambda)
-2\lambda (\Nc-1) ,
\end{align}
where $\{T_l\}_{i_{\alpha},n_{\alpha}}$ is {the sequence of  nonzero $T_l$ }
 in the $\alpha$th cluster-$n$.


The ground state is a uniform array of cluster-$n$'s with a common value of $n$ and dimers in between. 
This state is called the dimer-cluster-$n$ (DC$n$) state. 
The DC0 state is also called the dimer-monomer state. 
In the DC$n$ phase, we can take $T_{(n+1)p+q}=0$ and $\{T_l\}_{i,n}=\{T_l\}_n\equiv \{T_1...T_n\}$ with 
$i=(n+1)p+q+1$ for 
all integer $p$. 
The integer $q$ can be chosen arbitrarily within the range $0 \leq q \leq n$. 
Hence, this ground state has a spatial periodicity of $n+1$ with the $(n+1)$-fold spontaneous translational symmetry breakdown. 

The value of $n$ and the configuration $\{T_l\}_n$ are determined so as to minimize the total energy for each $\lambda$. The energy per unit cell of the lowest energy state among the DC$n$ states is given by  
\begin{align}
&\epsg(\lambda)={\rm min}_{n,\{T_l\}_{n}}\frac{1}{n+1} \left[\EG(\{T_l\}_{n},\lambda)-2\lambda\right].
\end{align}

The phase boundary $\lambda_{\rm c}(n;\{T_l\}_n,\{T'_l\}_n)$ between the DC$n$ phases with the $T$-spin configurations $\{T_l\}_n$ and $\{T'_l\}_n$ is given by
\begin{align}
&\lambda_{\rm c}(n;\{T_l\}_n,\{T'_l\}_n)\nonumber\\ 
&=\frac{2[\EHal_{\rm G}(\{T'_l\}_n)-\EHal_{\rm G}(\{T_l\}_n)]}{\sum_{l=1}^{n}[(T_{l}+1)T_l-(T'_{l}+1)T'_l]}.
\end{align}
If  $T_l\neq 0$ for all diagonal bonds, the ground-state energy of (\ref{ham}) per unit cell $\epsg(\infty,\{T_l\}, \lambda)$ can be expressed by the ground-state energy  $\epsgtilde(\{T_l\})$  of the infinite {\MAFH}  with the configuration $\{T_l\}$  as 
\begin{align}
\epsg(\infty,\{T_l\}, \lambda)=2\epsgtilde(\{T_l\}) +\frac{\lambda}{2}\overline{(T_{l}+1)T_l}-{2\lambda},
\end{align}
where 
\begin{align}
\overline{(T_{l}+1)T_l}=\lim_{L\rightarrow\infty}\frac{1}{L}\sum_{l=1}^{L}(T_{l}+1)T_l
\end{align}
is the average of $(T_{l}+1)T_l$ over the whole chain. 

In Ref. \citen{tks}, the following results are derived:
\begin{enumerate}
\item For $\lambda > 2$, $T_l=0$ or 1 for all $l$.  The ground state is the dimer-monomer (DC0) phase for $\lambda >3$, tetramer-dimer (DC1) phase for $2.660 \lesssim \lambda < 3$, heptamer-dimer (DC2) phase for  $ 2.583 \lesssim \lambda \lesssim 2.660$, DC3 phase for  $2.577  \lesssim \lambda \lesssim 2.583$, and DC$\infty$ phase for $2  < \lambda \lesssim 2.577$. In all these phases, $T_l=1$ for all diagonal bonds in cluster-$n$. In particular, the DC$\infty$ phase is the spin-1 Haldane phase.
\item For $\lambda <0$, $T_l=2$ for all $l$. Hence, the ground state is ferrimagnetic with magnetization $M=L$.
\end{enumerate}
However, the ground states for $0 < \lambda <2$ remained unresolved. In the present work, we determine the ground-state phases in this region.
\begin{figure} 
\centerline{\includegraphics[width=6cm]{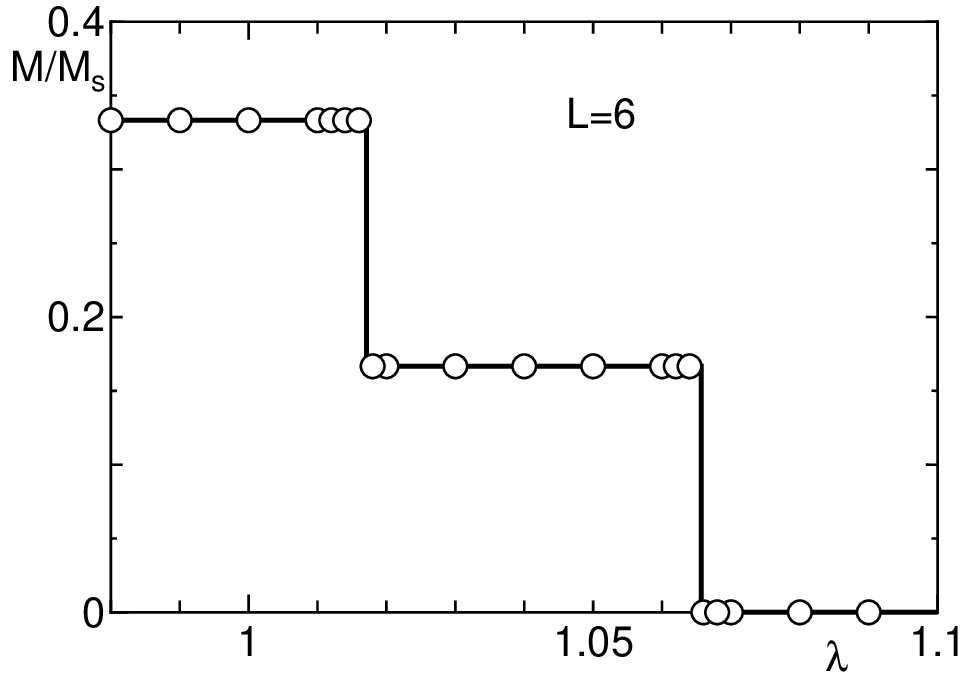}}
\caption{Ground-state magnetization for $L=6$.}
\label{fig:mag_s}
\end{figure}
\begin{figure} 
\centerline{\includegraphics[width=6cm]{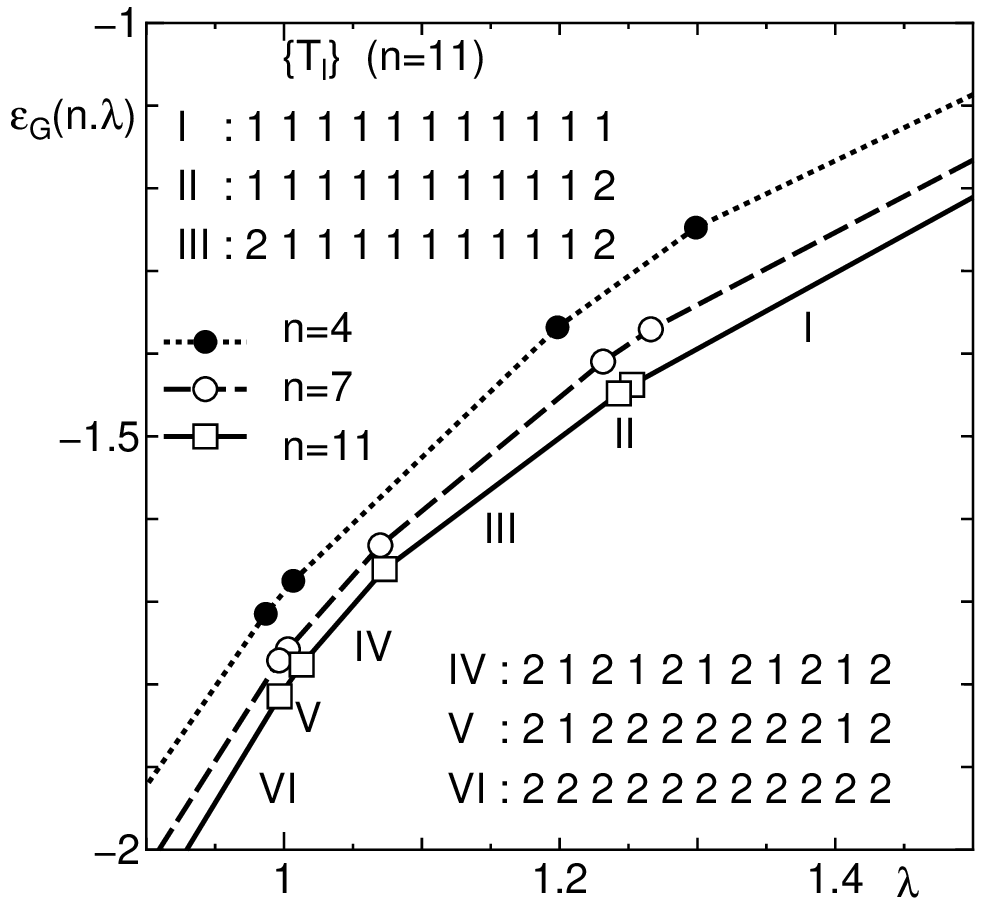}}
\caption{Lowest energies of the DC$n$ states.}
\label{fig:enesel}
\end{figure}

To find out the possible candidates of the ground-state configuration $\{T_l\}$ in the thermodynamic limit, we start with the numerical exact diagonalization of the whole chain with $L=6$ with periodic boundary condition. The total magnetization $M$ in the ground state 
is shown in Fig. \ref{fig:mag_s}. We find $M=\Ms/3$ for $\lambda \lesssim 1.0171$,  $M=\Ms/6$ for $1.0171 \lesssim \lambda \lesssim 1.0657$, and  $M=0$ for $\lambda \gtrsim 1.0657$, where $\Ms(=3L)$ is the saturated magnetization. No ground states with other values of magnetizations are found. 

To identify the ground-state phases in more detail, we estimate the ground-state energy assuming the DC$n$ state with $n \leq 11$ for all possible configurations $\{T_l\}_n$. The lowest energies of the DC$n$ states per site are plotted against $\lambda$ in Fig. \ref{fig:enesel} for several values of $n$ using the numerical exact diagonalization and finite-size DMRG methods. In the DMRG calculation, the number $m$ of  states kept in each subsystem is 240. The convergence with respect to $m$ is confirmed. 
The lowest energy of the DC$n$ state decreases with $n$. This suggests that the true ground state is the DC$\infty$ phase.

To obtain more insight into the configuration of $\{T_l\}$ in the thermodynamic limit, we identify the configuration $\{T_l\}_n$ in the DC$n$ state with finite $n$ in more detail. For example, in the DC11 phase, there are six regions with different lowest energy configurations as shown in Fig. \ref{fig:enesel}. However, the configurations in regions II and III differ from that in region I only locally. Similarly,  the configurations in regions V and VI differ from each other only locally. Hence, we speculate that the candidates of the configurations in the ground-state DC$\infty$ phases are the following three configurations.
\begin{enumerate}
\item  Haldane phase
\begin{align}
\{T_l\}&=\{\dot{1}\}.
\end{align}
\item Ferrimagnetic phase with $M=\Ms/6$ (F$_{1/6}$ phase)
\begin{align}
\{T_l\}&= \{\dot{1}\dot{2}\}.\label{f6}
\end{align}
\item Ferrimagnetic phase with $M=\Ms/3$ (F$_{1/3}$ phase)
\begin{align}
\{T_l\}&=\{\dot{2}\}.\label{f3}
\end{align}
\end{enumerate}
Here, $\{\dot{T_1}\cdots\dot{T}_l\}$ denotes the configuration consisting of a periodic array of $T_1\cdots T_l$ over the whole chain. The magnetizations in the ferrimagnetic phases F$_{1/3}$ and F$_{1/6}$ are stable against the variation of the parameter $\lambda$ within each phase. This implies that finite energy gaps open between the ferrimagnetic state and the excited states with different magnetizations. The presence of the gap in the F$_{1/6}$ and  F$_{1/3}$ phases is allowed by the theorem of Oshikawa {\it et al.}\cite{oya} and its extension by Nomura {\it et al.} \cite{nomura} assuming the configurations (\ref{f6}) and (\ref{f3}), respectively. In contrast to the F$_{1/3}$ phase, the configuration (\ref{f6}) of the F$_{1/6}$ phase breaks the translational symmetry and the space-inversion symmetry around the $S$-spin spontaneously, while the space-inversion symmetry around the $T$-spin is preserved.

Upper and lower bounds for the phase boundaries are obtained exactly by the assistance of numerical calculation, as follows. 
We first divide the whole Hamiltonian  (\ref{ham1}) as
\begin{align}
{\cal H}&={\cal H}_{\enu}
+{\cal \bar{H}}_{\enu} ,
\end{align}
where 
\begin{align}
{\cal H}_{\enu}&=\sum_{l=i}^{i+\enu-1} \left[\v{S}_{l}\v{T}_{l}+\v{T}_{l}\v{S}_{l+1}+ \frac{\lambda}{2}(\v{T}_l^2-4) \right] ,
\end{align}
and ${\cal \bar{H}}_{\enu}$ is the remainder 
 of ${\cal H}$. 
Consider an eigenstate of the whole diamond chain $\ket{\Phi(\{T_l\})}$ specified by $\{T_l\}$ that contains a sequence $\{T_l\}_{i,n}$ with $T_l\neq 0$ ($i \leq l \leq i+\enu-1$). The expectation value of the total Hamiltonian with respect to $\ket{\Phi(\{T_l\})}$ is expressed as
\begin{align}
\bra{\Phi(\{T_l\})}{\cal H}\ket{\Phi(\{T_l\})}&=\bra{\Phi(\{T_l\})}{\cal H}_{\enu}\ket{\Phi(\{T_l\})}\nonumber\\
&+\bra{\Phi(\{T_l\})}{\cal \bar{H}}_{\enu}\ket{\Phi(\{T_l\})}.
\end{align}
The following inequality holds in general,
\begin{align}
&\bra{\Phi(\{T_l\})}{\cal H}_{\enu}
\ket{\Phi(\{T_l\})}\geq\bra{\Phi_n(\{T_l\}_{i,n})}{\cal H}_{\enu}
\ket{\Phi_n(\{T_l\}_{i,n})}\nonumber\\
&=\EHal_{\rm G}(\{T_l\}_{i,n}\})+ \frac{\lambda}{2}\sum_{l=i}^{i+\enu-1}(T_{l}+1)T_l-{2\enu\lambda},
\end{align}
where $\ket{\Phi_n(\{T_l\}_{i,n})}$ is the lowest energy eigenstate of ${\cal H}_{\enu}
$ with the configuration $\{T_l\}_{i,n}$. Similarly,   we have an inequality,
\begin{align}
\bra{\Phi(\{T_l\})}{\cal \bar{H}}_{\enu}\ket{\Phi(\{T_l\})}&\geq\bar{E}_{\rm G},
\end{align}
where $\bar{E}_{{\rm G}}$ is the ground-state energy of ${\cal \bar{H}}_{\enu}$. 

Consider the lowest energy state $\ket{\Phi_0(\{T'_l\})}$ of the Hamiltonian  (\ref{ham1}) within the sector specified by the sequence $\{T'_l\}$ in which the subsequence  $\{T_l\}_{i,n}$ of $\{T_l\}$  is replaced by a sequence $\{T'_l\}_{i,n}=\{0T'_{i+1}...T'_{i+\enu-2}0\}$. Note that $T'_i=T'_{i+n-1}=0$ by definition. The zeros on both ends break the correlation  
between the spins in ${\cal H}_{\enu}
$ and those in ${\cal \bar{H}}_{\enu}$. 
 Then $\ket{\Phi_0(\{T'_l\})}$ can be expressed as 
\begin{align}
\ket{\Phi_0(\{T'_l\})}&=\ket{\Phi_n(\{T'_l\}_{i,n})}\otimes\ket{\bar{\Phi}},
\end{align}
where $\ket{\bar{\Phi}}$ is the ground state of ${\cal \bar{H}}_{\enu}$, which {satisfies}
\begin{align}
{\cal \bar{H}}_{\enu}\ket{\bar{\Phi}}&= \bar{E}_{{\rm G}} \ket{\bar{\Phi}}.
\end{align}
The expectation value of the total Hamiltonian with respect to $\ket{\Phi_0(\{T'_l\})}$ is {given by}
\begin{align}
\bra{\Phi_0(\{T'_l\})}{\cal H}\ket{\Phi_0(\{T'_l\})}&=\bra{\Phi_0(\{T'_l\})}{\cal H}_{\enu}\ket{\Phi_0(\{T'_l\})}\nonumber\\
&+\bra{\Phi_0(\{T'_l\})}{\cal \bar{H}}_{\enu}\ket{\Phi_0(\{T'_l\})},
\end{align}
where
\begin{align}
&\bra{\Phi_0(\{T'_l\})}{\cal H}_{\enu}\ket{\Phi_0(\{T'_l\})}=\bra{\Phi_n(\{T'_l\}_{i,n})}{\cal H}_{\enu}\ket{\Phi_n(\{T'_l\}_{i,n})}\nonumber\\
&=\EHal_{\rm G}(\{T'_l\}_{i+1,n-2})+ \frac{\lambda}{2}\sum_{l=i+1}^{i+\enu-2}(T'_{l}+1)T'_l-{2\enu\lambda},
\end{align}
\begin{align}
\bra{\Phi_0(\{T'_l\})}{\cal \bar{H}}_{\enu}\ket{\Phi_0(\{T'_l\})}&=\bar{E}_{{\rm G}}.
\end{align}
Hence, we have
\begin{align}
&\bra{\Phi(\{T_l\})}{\cal H}\ket{\Phi(\{T_l\})}-\bra{\Phi_0(\{T'_l\})}{\cal H}\ket{\Phi_0(\{T'_l\})}\nonumber\\
\geq
&\Delta \EHal_{i,\enu}(\{T_l\},\{T'_{l'}\}) + \frac{\lambda}{2}D_{i,\enu}(\{T_l\},\{T'_{l'}\}),\label{eq:udlim}
\end{align}
where 
\begin{align}
D_{i,\enu}(\{T_l\},\{T'_{l'}\})&=\sum_{l=i}^{i+\enu-1}(T_{l}+1)T_l- \sum_{l=i+1}^{i+\enu-2}(T'_{l}+1)T'_l,\\
\Delta \EHal_{i,\enu}(\{T_l\},\{T'_{l'}\}) &=\EHal_{\rm G}(\{T_l\}_{i,n})-\EHal_{\rm G}(\{T'_l\}_{i+1, {n-1}}). 
\end{align}

If the rhs of (\ref{eq:udlim}) is positive, then the lhs is also positive 
and $\ket{\Phi(\{T_l\})}$ is not a ground state. If $D_{i,\enu}(\{T_l\},\{T'_{l'}\})>0 (<0)$, this condition gives a lower (upper) bound of $\lambda$ for which $\ket{\Phi(\{T_l\})}$ is {\it not} a ground state.

\begin{figure} 
\centerline{\includegraphics[width=7cm]{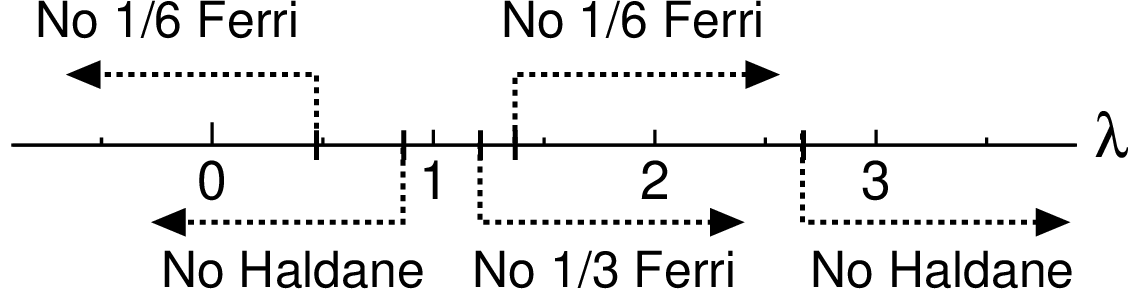}}
\caption{Upper and lower bounds for the phase boundaries estimated from the condition that the rhs of (\ref{eq:udlim}) is positive with $n=11$.}
\label{fig:bound}
\end{figure}

Using the finite-size DMRG results {with $m=240$} for 
$\enu=11$, we find that the Haldane phase is not the ground state for $\lambda  > 2.6717
$ and  $\lambda  <   0.86569$,  the F$_{1/6}$ phase is not the ground state for $\lambda > 1.3699$ and  $\lambda <  0.47135$,  and  the F$_{1/3}$ phase is not the ground state for $\lambda > 1.2125$, as summarized in Fig. \ref{fig:bound}. The convergence with respect to $m$ is confirmed.  
To guarantee the validity of the inequalities, the last digits of the lower bounds are rounded up and those of the upper bounds are rounded down. 
Actually, the phase boundaries are numerically estimated with high precision in the remainder of this paper. Nevertheless, it is worthwhile to determine the lower and upper bounds for the phase boundaries as above, since these are {\it exact} bounds and can be used to check if the numerically estimated values are appropriate.

 The ground-state energy of an infinite diamond chain per unit cell for each configuration is numerically estimated as follows: 
\begin{align}
&\epsg(\infty,\{\dot{1}\}, \lambda)=2\epsgtilde(\{\dot{1}\}) -{\lambda}\nonumber\\ 
&\simeq-2.8030-\lambda,
\end{align}
\begin{align}
&\epsg(\infty,\{\dot{1}\dot{2}\}, \lambda)=2\epsgtilde(\{\dot{1}\dot{2}\})\nonumber\\ 
&\simeq  -3.8757,
\end{align}
\begin{align}
&\epsg(\infty,\{\dot{2}\}, \lambda)=2\epsgtilde(\{\dot{2}\}) +{\lambda}\nonumber\\ 
&\simeq-4.8939+\lambda, 
\end{align}
where $\epsgtilde(\{\dot{T_1}\cdots\dot{T}_l\} )$ is the ground-state energy per spin of an infinite {\MAFH} with the configuration $\{\dot{T_1}\cdots\dot{T}_l\}$. 
 Among them,  $\epsgtilde(\{\dot{1}\})$ is the ground-state energy 
per spin for the antiferromagnetic Heisenberg chain with spin 1.\cite{white} 
We emplo{yed} the infinite-size DMRG method to evaluate $\epsgtilde(\{\dot{1}\dot{2}\})$  
and $\epsgtilde(\{\dot{2}\})$ with $m=480$. They are evaluated from the energy increment accompanied by an addition of a group  of spins specified by the sequence $\{12\}$ or $\{2\}$ at the center of the chain. The convergences with respect to $m$ and DMRG iteration are confirmed up to the digits presented above.

\begin{figure} 
\centerline{\includegraphics[width=7cm]{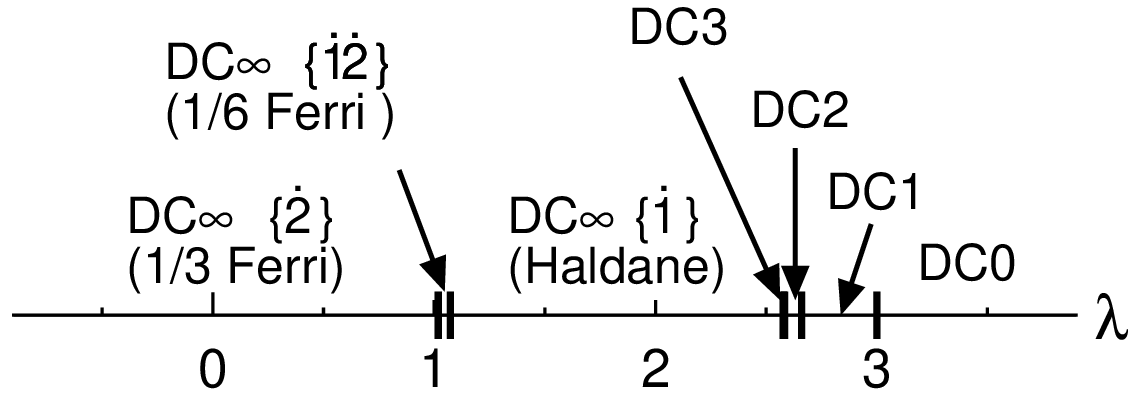}}
\caption{Ground state phase diagram.}
\label{fig:phase}
\end{figure}

The phase boundary between {the} Haldane phase and {the} F$_{1/6}$ phase is given by
\begin{align}
 {\lambda}_{\rm H F_{1/6}}=2\epsgtilde(\{\dot{1}\}) -2\epsgtilde(\{\dot{1}\dot{2}\})
\simeq 1.0727.
\end{align}
The phase boundary between the F$_{1/6}$ phase and the F$_{1/3}$ phase is given by
\begin{align}
 {\lambda}_{\rm F_{1/6}F_{1/3}}=2\epsgtilde(\{\dot{1}\dot{2}\})-2\epsgtilde(\{\dot{2}\})
\simeq 1.0182.
\end{align}
These phase boundaries are depicted in Fig. \ref{fig:phase} together with the other phase boundaries, which are given in Ref. \citen{tks}. 
All the phase boundaries are consistent with the upper and lower bounds shown 
 in {Fig. \ref{fig:bound}}

Our results are summarized as follows: The ground-state phase diagram of a spin-1 diamond chain is determined.  
Any eigenstate of this system can be expressed using the eigenstates of finite odd-length chains or an infinite chain with spins 1 and 2.  In addition to the various paramagnetic DC$n$ phases, the Haldane phase and the ferrimagnetic phases with 1/3 and 1/6 of the saturated magnetization are found. Among them, the ferrimagnetic phase with 1/6 magnetization is accompanied by the spontaneous  breakdown of translational symmetry and the space-inversion symmetry around the $S$-spin that was not found in previous works on diamond chains while the space-inversion symmetry around the $T$-spin is preserved. Exact upper and lower bounds for the phase boundaries are also obtained.

In real materials, the ideal diamond chain is hardly realized. 
Hence, it is necessary to investigate the effect of perturbations such as lattice distortion, anisotropy, and interchain coupling. Among them, the lattice distortion is expected to transform the paramagnetic DC$n$ phases into various types of Haldane phases and ferrimagnetic phases 
as discussed in  
the mixed diamond chain in Ref. \citen{hts3}. 
The detailed results will be reported elsewhere. 

In the case of larger spin magnitude, a larger variety of phases are expected to appear. These are left for future studies.

\acknowledgments

The authors are grateful to H. Kikuchi for explaining the experimental results of his group prior to publication\cite{kuni,kiku1} and for drawing our attention to Ref. \citen{guillou}. The numerical diagonalization program is based on the package TITPACK ver. 2 coded by H. Nishimori.  Part of the numerical computation in this work was carried out using the facilities of the Supercomputer Center, Institute for Solid State Physics, University of Tokyo, and   Yukawa Institute Computer Facility in Kyoto University. This work is  supported by JSPS KAKENHI Grant Numbers JP25400389 and JP26400411.

\end{document}